\documentstyle[times,pramana,epsfig,floats]{ias}

\def\PRC{{\em Phys. Rev.} {\bf C}}
\def\NPA{{\em Nucl. Phys.} {\bf A}}
\def\PLB{{\em Phys. Lett.} {\bf B}}
\def\PRL{{\em Phys. Rev. Lett.} }

\newcommand{\be}{\begin{equation}}
\newcommand{\ee}{\end{equation}}
\newcommand{\bea}{\begin{eqnarray}}
\newcommand{\eea}{\end{eqnarray}}

\begin{document}
\title{ Hadrons in Medium}
\author{U. Mosel}
\address{
Institut fuer Theoretische Physik, Universitaet Giessen\\
D-35392 Giessen, Germany} \maketitle \abstract{ After a short
motivation I outline a consistent treatment of hadronic spectral
functions based on transport theory. As examples I discuss nucleon
spectral functions and the observable effects of changes of the
properties of vector mesons inside nuclei.}


\section{Introduction}

Studies of in-medium properties of hadrons are driven by a number of
- partly connected - motivations. A first motivation for the study
of hadronic in-medium properties is provided by our interest in
understanding the structure of large dense systems, such as the
interior of stars. This structure obviously depends on the
composition of stellar matter and its interactions (for a recent
review see \cite{Weber}). On a smaller scale, this also holds for
the structure of nuclei which is influenced by the properties of
their building blocks, the baryons, and that of the exchange mesons
that provide the binding.

The second motivation is based on the expectation that changes of
hadronic properties in medium can be precursor phenomena to changes
in the phase structure of nuclear matter. Here the transition to the
chirally symmetric phase, that exhibits manifestly the symmetries of
the underlying theory of strong interactions, i.e. QCD, is of
particular interest. Present day's ultrarelativic heavy-ion
collisions explore this phase boundary in the limit of high
temperatures ($T \approx$ 170 MeV) and low densities. The other
limit (low temperatures and high densities) is harder to reach,
although the older AGS heavy-ion experiments and the planned CBM
experiment at the new FAIR facility \cite{FAIR} may yield insight
into this area. However, even in these experiments the temperatures
reached are still sizeable ($T \approx $ 120 MeV). At very low
temperatures the only feasible method seems to be the exploration of
the hadronic structure inside ordinary nuclei, at the prize of a low
density. Here the temperature is $T=0$ and the density at most
equals the equilibrium density of nuclear matter, $\rho_0$. It is
thus of great interest to explore if such low densities can already
give precursor signals for chiral symmetry restoration.

\subsection{Phenomenology}
 That hadrons can indeed change their properties and
couplings in the nuclear medium has been well known to nuclear
physicists since the days of the Delta-hole model that dealt with
the changes of the properties of the pion and Delta-resonance inside
nuclei \cite{Ericsson-Weise}. Due to the predominant $p$-wave
interaction of pions with nucleons one observes here a lowering of
the pion branch with increasing pion-momentum and nucleon-density.
This effect can be seen in optical model analyses of pion scattering
on nuclei, but the absorptive part of the $\pi$-nucleus interaction
limits the sensitivity to small densities. More recently,
experiments at the FSR at GSI have shown that also the mass of a
pion at rest in the nuclear medium differs from its value in vacuum
\cite{Kienle}. This is interesting since there are also recent
experiments \cite{TAPSsigma} that look for in-medium changes of the
$\sigma$ meson, the chiral partner of the pion. Any comparison of
scalar and pseudoscalar strength could thus give information about
the degree of chiral symmetry restoration in nuclear matter.

In addition, experiments for charged kaon production at GSI
\cite{KAOS} have given some evidence for the theoretically predicted
lowering of the $K^-$ mass in medium and the (weaker) rising of the
$K^+$ mass. State-of-the-art calculations of the in-medium
properties of kaons have shown that the usual quasi-particle
approximation for these particles is no longer justified inside
nuclear matter where they acquire a broad spectral function
\cite{Lutz,Tolos}.

At higher energies, at the CERN SPS and most recently at the
Brookhaven RHIC, in-medium changes of vector mesons have found
increased interest. The interest in vector meson production stems
from the fact that these mesons couple strongly to the photon so
that electromagnetic signals can yield information about properties
of hadrons deeply embedded into nuclear matter. Indeed, the CERES
experiment \cite{CERES} has found a considerable excess of dileptons
in an invariant mass range from $\approx 300$ MeV to $\approx 700$
MeV as compared to expectations based on the assumption of freely
radiating mesons.

This result has found an explanation in terms of a shift of the
$\rho$ meson spectral function down to lower masses, as expected
from theory (see, e.g., \cite{Postneu,Peters,Post,Wambach}).
However, the actual reason for the observed dilepton excess is far
from clear. This is so partly because any signals from heavy-ion
collisions are time-integrals over the often quite complex collision
history with very different densities and temperatures.

I have therefore already some years ago proposed to look for the
theoretically predicted changes of vector meson properties inside
the nuclear medium in reactions on normal nuclei with more
microscopic probes \cite{Hirschegg97,Hirschegg}. Of course, the
nuclear density felt by the vector mesons in such experiments lies
much below the equilibrium density of nuclear matter, $\rho_0$, so
that naively any density-dependent effects are expected to be much
smaller than in heavy-ion reactions.

On the other hand, there is a big advantage to these experiments:
they proceed with the spectator matter being close to its
equilibrium state. This is essential because all theoretical
predictions of in-medium properties of hadrons are based on an
equilibrium model in which the hadron (vector meson) under
investigation is embedded in cold nuclear matter in equilibrium and
with infinite extension. The properties so calculated are then, in a
second step, being locally inserted into a time-dependent event
simulation.  In actual experiments these hadrons are observed
through their decay products and these have to travel through the
surrounding nuclear matter to the detectors. Except for the case of
electromagnetic signals (photons, dileptons) this is connected with
often sizeable final state interactions (FSI) that have to be
treated as realistically as possible. For a long period the Glauber
approximation which allows only for absorptive processes along a
straight-line path has been the method of choice in theories of
photonuclear reactions on nuclei. This may be sufficient if one is
only interested in total yields of strongly absorbed particles.
However, it is clearly insufficient when one aims at, for example,
reconstructing the spectral function of a hadron inside matter
through its decay products. Rescattering and sidefeeding through
coupled channel effects can affect the final result so that a
realistic description of such effects is absolutely mandatory
\cite{FalterShad}.

\section{Theory}

\subsection{Chiral Symmetry}

A large part of the current interest in in-medium properties of
hadrons comes from the hope to learn something about quarks in
nuclei. More specifically, one hopes to see precursors of a
restoration of the original symmetries of the theory of strong
interactions, i.e.\ QCD, which are spontaneously broken in our
world. In \cite{MoselPuri} and in particular in \cite{Moselbuch} I
have discussed this point at some length. Here I just state that the
so-called NJL model that is manifestly chirally invariant indeed
leads to a linear relationship between the fermion mass and the
chiral condensate, the order parameter of chiral symmetry breaking.
In this model this condensate in turn drops with baryon density and
temperature.

In the NJL model the dropping of the chiral condensate with density
and/or temperature directly causes a drop of the mass because both
are linearly proportional. This is no longer the case in complex
hadronic systems. How the drop of the scalar condensate there
translates into observable hadron masses is not uniquely prescribed.
The only rigorous connection is given by the QCD sum rules that
relate an integral over the hadronic spectral function to a sum over
combinations of quark- and gluon-condensates with powers of $1/Q^2$
\cite{Peskin-Schroeder}.

Since the spectral function appears under an integral the
information obtained is, however, not very specific. However,
Leupold et al. have shown \cite{Leupold,LeupoldMosel} that the QCDSR
provides important constraints for the hadronic spectral functions
in medium, but it does not fix them. Recently Kaempfer et al have
turned this argument around by pointing out that measuring an
in-medium spectral function of the $\omega$ meson could help to
determine the density dependence of the chiral condensate
\cite{Kaempfer}.

\subsection{QCD Sum Rules and Collisional Broadening}

Since QCD sum rules do not fix the hadronic properties models are
needed for the hadronic interactions. The quantitatively reliable
ones can at present be based only on 'classical' hadrons and their
interactions. Indeed, in lowest order in the density the mass and
width of an interacting hadron in nuclear matter at zero temperature
and vector density $\rho_v$ are given by (for a meson, for example)
\begin{eqnarray}     \label{trho}
{m^*}^2 = m^2 - 4 \pi \Re f_{m N}(q_0,\theta = 0)\, \rho_v
\nonumber \\
m^* \Gamma^* = m \Gamma^0 -  4 \pi \Im f_{mN}(q_0,\theta = 0)\,
\rho_v ~.
\end{eqnarray}
Here $f_{mN}(q_0,\theta = 0)$ is the forward scattering amplitude
for a meson with energy $q_0$ on a nucleon. The width $\Gamma^0$
denotes the free decay width of the particle. For the imaginary part
this is nothing other than the classical relation $\Gamma^* -
\Gamma^0 = v \sigma \rho_v$ for the collision width, where $\sigma$
is the total cross section. This can easily be seen by using the
optical theorem.

Actually evaluating mass and width from (\ref{trho}) requires
knowledge of the scattering amplitude which can only be obtained
from very detailed analyses of experiments. The $s$-channel
contributions to this scattering amplitude are determined by the
properties of nucleon resonances and these are often not very well
known yet. Here, resonance physics meets in-medium physics.

Unfortunately it is not a-priori known up to which densities the
low-density expansion (\ref{trho}) is useful. Post et al.
\cite{Postneu} have recently investigated this question in a
coupled-channel calculation of selfenergies. Their analysis
comprises pions, $\eta$-mesons and $\rho$-mesons as well as all
baryon resonances with a sizeable coupling to any of these mesons.
The authors of \cite{Postneu} find that already for densities less
than $0.5 \rho_0$ the linear scaling of the selfenergies inherent
in (\ref{trho}) is badly violated for the $\rho$ and the $\pi$
mesons, whereas it is a reasonable approximation for the $\eta$
meson. Reasons for this deviation from linearity are Fermi-motion,
Pauli-blocking, selfconsistency and short-range correlations. For
different mesons different sources of the discrepancy prevail: for
the $\rho$ and $\eta$ mesons the iterations act against the
low-density theorem by inducing a strong broadening for the
$D_{13}(1520)$ and a slightly repulsive mass shift for the
$S_{11}(1535)$ nucleon resonances to which the $\rho$ and the
$\eta$ meson, respectively, couple. The investigation of in-medium
properties of mesons, for example, thus involves at the same time
the study of in-medium properties of nucleon resonances and is
thus a coupled-channel problem.

\subsection{Coupled Channel Treatment of Incoherent Particle
Production}\label{sec:CoupledChannel}

In order to avoid the intrinsic difficulties connected with using
equilibrium hadronic properties in a non-equilibrium situation such
as a heavy-ion reaction, we have looked for possible effects in
reactions that proceed closer to equilibrium, i.e. reactions of
elementary probes such as protons, pions, and photons on nuclei. The
densities probed in such reactions are always $\le \rho_0$, with
most of the nucleons actually being at about $0.5 \rho_0$. On the
other hand, the target is stationary and the reaction proceeds much
closer to (cold) equilibrium than in a relativistic heavy-ion
collision. If any observable effects of in-medium changes of
hadronic properties survive, even though the densities probed are
always $\le \rho_0$, then the study of hadronic in-medium properties
in reactions with elementary probes on nuclei provides an essential
baseline for in-medium effects in hot nuclear matter probed in
ultra-relativistic heavy-ion collisions.

With the aim of exploring this possibility we have over the last few
years undertaken a number of calculations for proton-
\cite{Bratprot}, pion- \cite{Weidmann,Effepi} and photon-
\cite{Effephot} induced reactions. All of them have one feature in
common: they treat the final state incoherently in a coupled channel
transport calculation that allows for elastic and inelastic
scattering of, particle production by and absorption of the produced
hadrons. A new feature of these calculations is that hadrons with
their correct spectral functions can actually be produced and
transported consistently. This is quite an advantage over earlier
treatments \cite{Brat-Cass,Fuchs} in which the mesons were always
produced and transported with their pole mass and their spectral
function was later on folded in only for their decay. The method is
summarized in the following section, more details can be found in
\cite{Effephot}.

We separate the photonuclear reaction into three steps. First, we
determine the amount of shadowing for the incoming photon; this
obviously depends on its momentum transfer $Q^2$. Second, the
primary particle is produced and third, the produced particles are
propagated through the nuclear medium until they leave the
nucleus.

\paragraph{Shadowing.} Photonuclear reactions show shadowing in the
entrance channel, for real photons from an energy of about 1 GeV
on upwards \cite{Bianchi}. This shadowing is due to a coherent
superposition of bare photon and vector meson components in the
incoming photon and is handled here by means of a Glauber multiple
scattering model \cite{FalterShad}. In this way we obtain for each
value of virtuality $Q^2$ and energy $\nu$ of the photon a spatial
distribution for the probability that the incoming photon reaches
a given point; for details see \cite{FalterShad,Effe,Falterinc}.

\paragraph{Initial Production.}
The initial particle production is handled differently depending
on the invariant mass $W = \sqrt{s}$ of the excited state of the
nucleon. If $W < 2$ GeV, we invoke a nucleon resonance model that
has been adjusted to nuclear data on resonance-driven particle
production \cite{Effephot}. If $W > 2$ GeV the particle yield is
calculated with standard codes developed for high energy nuclear
reactions, i.e.\ FRITIOF or PYTHIA; details are given in
\cite{Falterneu}. We have made efforts to ensure a smooth
transition of cross sections in the transition from resonance
physics to DIS.

\paragraph{Groundstate Correlations and Final State Interactions.}
The groundstate correlations and the final state interactions are
consistently described by a semiclassical coupled channel transport
theory that had originally been developed for the description of
heavy-ion collisions and has since then been applied to various more
elementary reactions on nuclei with protons, pions and photons in
the entrance channel.

In this method the spectral phase space distributions of all
particles involved are propagated in time, from the initial first
contact of the photon with the nucleus all the way to the final
hadrons leaving the nuclear volume on their way to the detector. The
spectral phase space distributions $F_h(\vec{r},\vec{p},\mu,t)$ give
at each moment of time and for each particle class $h$ the
probability to find a particle of that class with a (possibly
off-shell) mass $\mu$ and momentum $\vec{p}$ at position $\vec{r}$.
Its time-development is determined by the BUU equation
\be     \label{BUU}
(\frac{\partial}{\partial t} + \frac{\partial H_h}{\partial \vec{p}}
\frac{\partial}{\partial \vec{r}} - \frac{\partial H_h}{\partial
\vec{r}} \frac{\partial}{\partial \vec{p}})F_h=G_h a_h - L_h F_h.
\ee
Here $H_h$ gives the energy of the hadron $h$ that is being
transported; it contains the mass, the selfenergy (mean field) of
the particle and a term that drives an off-shell particle back to
its mass shell. The terms on the lhs of (\ref{BUU}) are the
so-called \emph{drift terms} since they describe the independent
transport of each hadron class $h$. The terms on the rhs of
(\ref{BUU}) are the \emph{collision terms}; they describe both
elastic and inelastic collisions between the hadrons. Here the term
\emph{inelastic collisions} includes those collisions that either
lead to particle production or particle absorption. The former is
described by the \emph{gain term} $G_h a_h$ on the rhs in
(\ref{BUU}), the latter process (absorption) by the \emph{loss term}
$L_h F_h$. Note that the gain term is proportional to the spectral
function $a$ of the particle being produced (see discussion below),
thus allowing for production of off-shell particles. On the
contrary, the loss term is proportional to the spectral phase space
distribution itself: the more particles there are the more can be
absorbed. The terms $G_h$ and $L_h$ on the rhs give the actual
strength of the gain and loss terms, respectively. They have the
form of Born-approximation collision integrals and take the
Pauli-principle into account. The free collision rates themselves
are taken from experiment or are calculated \cite{Effephot}.

The collision therm on the lhs of (\ref{BUU}) is responsible for the
collision broadening that all particles experience when they are
embedded in a dense medium. Collisions either change energy and
momentum of the particles are absorb them alltogether. Both
processes contribute to collisional broadening.

The detailed structure of the gain and loss terms can be obtained
from quantum transport theory \cite{KadBaym,BotMal}. To see this I
start by summarizing briefly the known fundamental relations of
quantum transport theory for the description of non-stationary
processes in an interacting quantum system, following closely the
presentation in \cite{Lehrspect}. In quantum transport theory the
non-stationary processes which introduce a coupling between causal
and anti-causal single particle propagation are described by the
one-particle correlation functions
\begin{equation}
g^>(1,1') = -i \langle \Psi(1)\Psi^\dagger(1')\rangle \qquad
g^>(1,1') = i \langle \Psi^\dagger(1')\Psi(1)\rangle~.
\end{equation}
where $\Psi$ are the nucleon field operators in Heisenberg
representation. Correspondingly, in an interacting quantum system
the single particle self-energy operator includes correlation
self-energies $\Sigma^{<>}$ which couple particle and hole degrees
of freedom \cite{KadBaym,BotMal}. Clearly, $g^{<>}$ and
$\Sigma^{<>}$  are closely related. The wanted relation is obtained
from transport theory. After a Fourier transformation to
energy-momentum representation the corresponding self-energies are
found as \cite{KadBaym}
\begin{eqnarray}
\Sigma^>(\omega,p) &=& g \int \frac{d^3p_2\,d\omega_2}{(2 \pi)^4}
\frac{d^3p_3\,d\omega_3}{(2 \pi)^4} \frac{d^3p_4\,d\omega_4}{(2
\pi)^4} \left(2\pi\right)^4 \delta^4(p + p_2 - p_3 - p_4)
|\bar{\mathcal{M}}|^2 \nonumber \\
& & \mbox{} \times g^<(\omega_2,p_2) g^>(\omega_3,p3)
g^>(\omega_4,p_4) \nonumber \\[2mm]
\Sigma^<(\omega,p) &=& g \int \frac{d^3p_2\,d\omega_2}{(2 \pi)^4}
\frac{d^3p_3\,d\omega_3}{(2 \pi)^4} \frac{d^3p_4\,d\omega_4}{(2
\pi)^4} \left(2\pi\right)^4 \delta^4(p + p_2 - p_3 - p_4)
|\bar{\mathcal{M}}|^2 \nonumber \\
& & \mbox{} \times g^>(\omega_2,p_2) g^<(\omega_3,p_3)
g^<(\omega_4,p_4)~.
\end{eqnarray}
Here, g = 4 is the spin-isospin degeneracy factor and $|\bar{
\mathcal{M}}|^2$ denotes the square of the in-medium nucleon-nucleon
scattering amplitude, averaged over spin and isospin of the incoming
nucleons and summed over spin and isospin of the outgoing nucleons.
Note that energy $\omega$ and three-momentum $p$ are not related by
a dispersion relation thus allowing for a description of the full
off-shell behavior of the self-energies.

Since both $g^{<>}$ and $\Sigma^{<>}$ describe the correlation
dynamics, the spectral function can be obtained from either of the
two quantities as the difference over the cut along the energy real
axis. In terms of the correlation propagators, the spectral density
is defined by
\begin{equation}
a(\omega,p) = i \left( g^>(\omega,p) - g<(\omega,p) \right) ~.
\end{equation}
Non-relativistically, the single particle spectral function is
explicitly found as
\begin{equation}
a(\omega,p) = \frac{\Gamma(\omega,p)}{\left( \omega -
\frac{p^2}{2m_N} - \Re{\Sigma(\omega,p)}\right)^2 +
\frac{1}{4}\Gamma^2(\omega,p)} ~,
\end{equation}
including the particle and hole nucleon self-energy $\Sigma$. The
width $\Gamma$ is given by the imaginary part of the retarded
self-energy
\begin{equation}
\Gamma(\omega, p) = 2\Im \Sigma(\omega, p) = i(\Sigma^>(\omega, p) -
\Sigma^<(\omega, p)).
\end{equation}
In the limiting case of vanishing correlations, i.e. $\Im \Sigma \to
0$, the usual deltalike quasi-particle spectral function is
recovered.

The correlation functions $g^{<>}$ can be re-written in terms of the
phasespace distribution function $f(\omega, p)$
\begin{eqnarray}
g^>(\omega,p) &=& - i a(\omega,p)(1 - f(\omega,p) \nonumber \\
g^<(\omega,p) &=& i a(\omega,p)f(\omega,p) ~.
\end{eqnarray}
For a system at $T = 0$, $f$ reduces to
\begin{equation}
f(\omega, p) = \Theta(\omega_F - \omega)~.
\end{equation}
As a result,we obtain for the self-energies the conditions
\begin{eqnarray}
\Sigma^>(\omega,p) = 0 &\hspace{5mm} & \Gamma(\omega,p) = -i
\Sigma^<(\omega,p)
\qquad \mbox{for} \quad \omega \le \omega_F \nonumber \\
\Sigma^<(\omega,p) = 0 & & \Gamma(\omega,p) = i \Sigma^>(\omega,p)
\qquad \mbox{for} \quad \omega \ge \omega_F ~.
\end{eqnarray}

The correlation function $-i g^<(\omega,p)$ is nothing else than the
Fourier-transform of the spectral phase space density $F_h$ in Eq.\
(\ref{BUU}), with the variable $\mu$ in (\ref{BUU}) being an
'off-shell mass' $\mu = \sqrt{\omega^2 - p^2}$. Also the gain and
loss terms in (\ref{BUU}) can be expressed in terms of the
correlations functions defined here. As a result, we have
\begin{equation}
G = i \Sigma^< (1 - f) \qquad L = i \Sigma^> ~.
\end{equation}

Due to the dependence of the width and the single particle spectral
function upon each other the calculation of $a(\omega, p)$ requires
a self-consistent treatment. Therefore, the transport theoretical
approach leads to single particle propagators including correlation
self-energies to all orders. The collision rates embedded in $G$ and
$L$ determine the collisional broadening of the particles involved
and thus their spectral function $a$. The widths of the particles,
resonances or mesons, thus evolve in time away from their vacuum
values. In addition, broad particles can be produced off their peak
mass and then propagated. The extra 'potential' in $H$ already
mentioned ensures that all particles are being driven back to their
mass-shell when they leave the nucleus. The actual method used is
described in \cite{Effephot}. It is based on an analysis of the
Kadanoff-Baym equation that has led to practical schemes for the
propagation of off-shell particles
\cite{Leupoldoffshell,JuchemCassing}´. The possibility to transport
off-shell particles represents a major breakthrough in this field.
For further details of the model see Ref. \cite{Effephot} and
\cite{Falterneu} and references therein.

\section{In-medium Hadrons -- Observables}

\subsection{Nucleon Spectral Functions}
\begin{figure}[ht]

\centerline{\epsfig{file=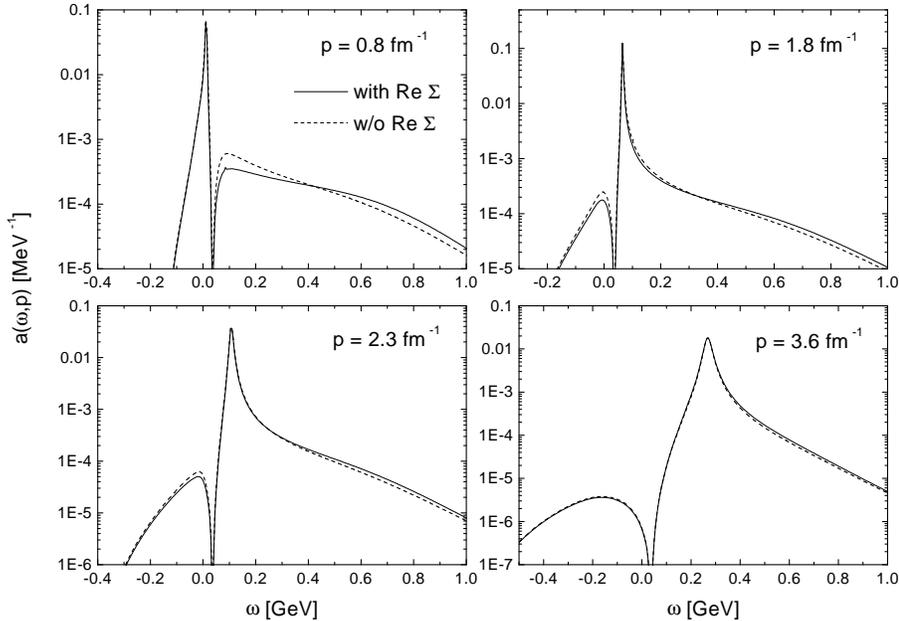,scale=1.0}} \caption{Nucleon
spectral function at momenta below and above the Fermi momentum
($1.37 fm^{-1}$) (from \protect\cite{Lehrspect}).}\label{fig:spektr}
\end{figure}

Even the elementary building blocks of nuclei, the nucleons, are
medium-modified inside nuclei. Due to correlations they pick up a
width and a mass-shift, the latter on top of that caused by the long
range mean field. This becomes evident by considering the case of
elastic collisions only in Eq.\ (\ref{BUU}). Nucleons moving in a
mean field can still collide, if the final states are above the
Fermi-surface thus acquiring a collisional width. This is
illustrated in Fig.\ \ref{fig:spektr}.

The nucleon spectral functions in Fig. \ref{fig:spektr} show the
distinct quasiparticle peak at the energy $\omega = \sqrt{p^2 +
m^2}$, broadened by collisions with the other nucleons. The width
gets larger with increasing momentum due to the phase space opening
up at the higher momenta. In addition to the quasiparticle peak the
spectral functions exhibit a distinct zero at the Fermi energy,
caused by the Pauli principle.

Observable effects of this broadening of nucleon spectral functions
in medium are well known: they show up in $(e,e'p)$ reactions on
nuclei where an analysis of the momentum and energy of the outgoing
proton allows to determine its spectral function inside the nucleus.
The method has recently been used to calculate the spectral function
of nucleons also at finite temperatures and higher densities
\cite{Froemelspect}, as well as for isospin-asymmetric nuclear
matter \cite{Konrad}

\subsection{\it $\omega$ Production}

Many of the early studies of hadronic properties in medium
concentrated on the $\rho$-meson \cite{Peters,Klingl}, partly
because of its possible significance for an interpretation of the
CERES experiment. It is clear by now, however, that the dominant
effect on the in-medium properties of the $\rho$-meson is
collisional broadening that overshadows any possible mass shifts
\cite{Postneu} and is thus experimentally hard to observe. The
emphasis has, therefore, shifted to the $\omega$ meson. An
experiment measuring the $A(\gamma,\omega \rightarrow
\pi^0\gamma')X$ reaction is presently being analyzed by the
TAPS/Crystal Barrel collaborations at ELSA \cite{Messch}. The
varying theoretical predictions for the $\omega$ mass (640-765 MeV)
\cite{Klingl} and width (up to 50 MeV) \cite{Weidmann,Friman} in
nuclear matter at rest encourage the use of such an exclusive probe
to learn about the $\omega$ spectral distribution in nuclei.

\begin{figure}[h]

\centerline{\epsfig{file=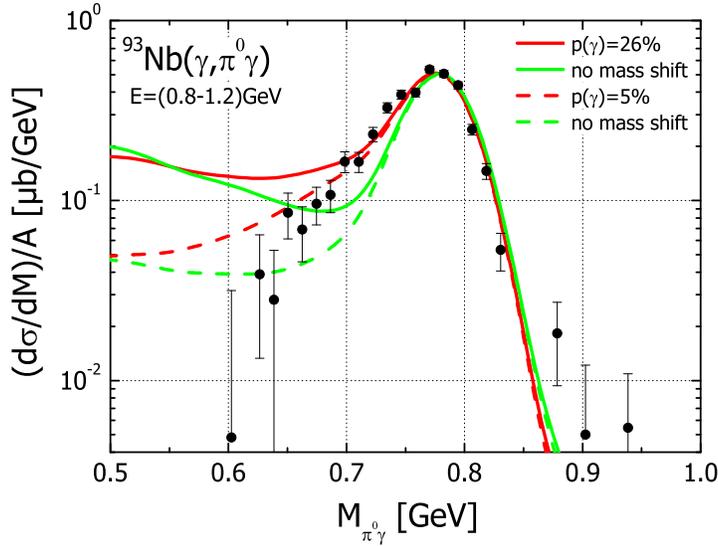,scale=1.0}}

\caption{Mass differential cross section for $\pi^0\gamma$
photoproduction off $^{93}$Nb. Shown are results both with and
without a mass-shift as explained in \protect\cite{MuehlOm}. The
quantity $p$ gives the escape probability for one of the four
photons in the $2\pi^0$ channel. The two solid curves give results
of calculations with p = 26 \% with and without mass-shift, the two
dashed curves give the same for p = 5\% (from
\protect\cite{Muehlichpriv}\label{omega}}
\end{figure}

Simulations have been performed at 1.2 GeV and 2.5 GeV photon
energy, which cover the accessible energies of the TAPS/Crystal
Barrel experiment. After reducing the combinatorial and rescattering
background by applying kinematic cuts on the outgoing particles, we
have obtained rather clear observable signals for an assumed
dropping of the $\omega$ mass inside nuclei \cite{MuehlOm}.
Therefore, in this case it should be possible to disentangle the
collisional broadening from a dropping mass.

Our calculations represent complete event simulations. It is,
therefore, possible to calculate these background contributions and
to take experimental acceptance effects into account. An example is
shown in Fig.~\ref{omega} which shows the effects of a possible
misidentification of the $\omega$ meson. This misidentification can
come about through the $2 \pi^0 \rightarrow 4 \gamma$ channel if one
of the four photons escapes detection and the remaining three
photons are identified as stemming from the $\pi^0 \gamma
\rightarrow 3 \gamma$ decay channel of the $\omega$-meson. The
calculations show that the misidentification does not affect the
low-mass side of the omega spectral function.

Fig.\ \ref{omega} shows a good agreement between the data of the
TAPS/CB@ELSA collaboration \cite{TrnkaPRL} for a photon escape
probability of 5 \% and a mass shift $m_\omega = m_\omega^0 - 0.18
\,\rho/\rho_0$. In \cite{MuehlOm} we have also discussed the
momentum-dependence of the $\omega$-selfenergy in medium and have
pointed out that this could be accessible through measurements which
gate on different three-momenta of the $\omega$ decay products.

\subsection{\it Dilepton Production}

Dileptons, i.e.\ electron-positron pairs, in the outgoing channel
are an ideal probe for in-medium properties of hadrons since they
experience no strong final state interaction. A first experiment to
look for these dileptons in heavy-ion reactions was the DLS
experiment at the BEVALAC in Berkeley \cite{DLS}. Later on, and in a
higher energy regime, the CERES experiment has received a lot of
attention for its observation of an excess of dileptons with
invariant masses below those of the lightest vector mesons
\cite{CERES}.
\begin{figure}[h]
\begin{center}
\begin{minipage}[t]{8 cm}
\centerline{\epsfig{file=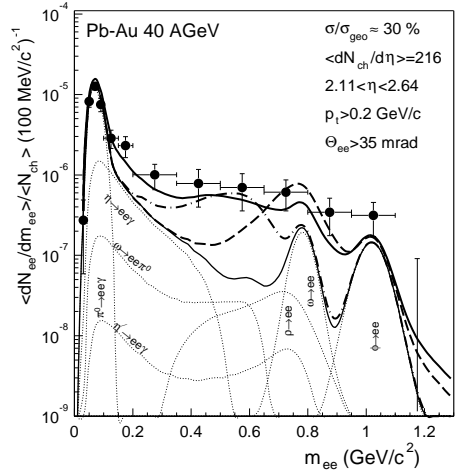,scale=0.6}}
\end{minipage}
\begin{minipage}[t]{16.5 cm}
\caption{Invariant  dilepton mass spectrum
 obtained with the CERES experiment in Pb + Au collisions at 40
 AGeV (from \protect\cite{CERES}). The thin curves give the contributions
 of individual hadronic sources to the total dilepton yield, the
 fat solid (modified spectral function) and the dash-dotted
 (dropping mass only) curves give the results of calculations
 \protect\cite{Rapp} employing an in-medium modified spectral function of the vector
 mesons.} \label{CERES}
\end{minipage}
\end{center}
\end{figure}
Explanations of this excess have focused on a change of in-medium
properties of these vector mesons in dense nuclear matter (see e.g.\
\cite{Cassingdil,RappWam}). The radiating sources can be nicely seen
in Fig.~\ref{CERES} that shows the dilepton spectrum obtained in a
low-energy run at 40 AGeV together with the elementary sources of
dilepton radiation.

The figure exhibits clearly the rather strong contributions of the
vector mesons -- both direct and through their Dalitz decay -- at
invariant masses above about 500 MeV. The strong amplification of
the dilepton rate at small invariant masses $M$ caused by the photon
propagator, which contributes $\sim 1/M^4$ to the cross section,
leads to a strong sensitivity to changes of the spectral function at
small masses. Therefore, the excess observed in the CERES experiment
can be explained by such changes as has been shown by various
authors (see e.g.\ \cite{Brat-Cass} for a review of such
calculations).

In view of the uncertainties in interpreting these results discussed
earlier we have studied the dilepton photo-production in reactions
on nuclear targets. Looking for in-medium changes in such a reaction
is not \emph{a priori} hopeless: Even in relativistic heavy-ion
reactions only about 1/2 of all dileptons come from densities larger
than $2 \rho_0$ \cite{Brat-Cass}. In these reactions the
pion-density gets quite large in the late stages of the collision.
Correspondingly many $\rho$ mesons are formed (through $\pi + \pi
\to \rho$) late in the collision, where the baryonic matter expands
and its density becomes low again.

\begin{figure}[ht]
 \centerline{\epsfig{file=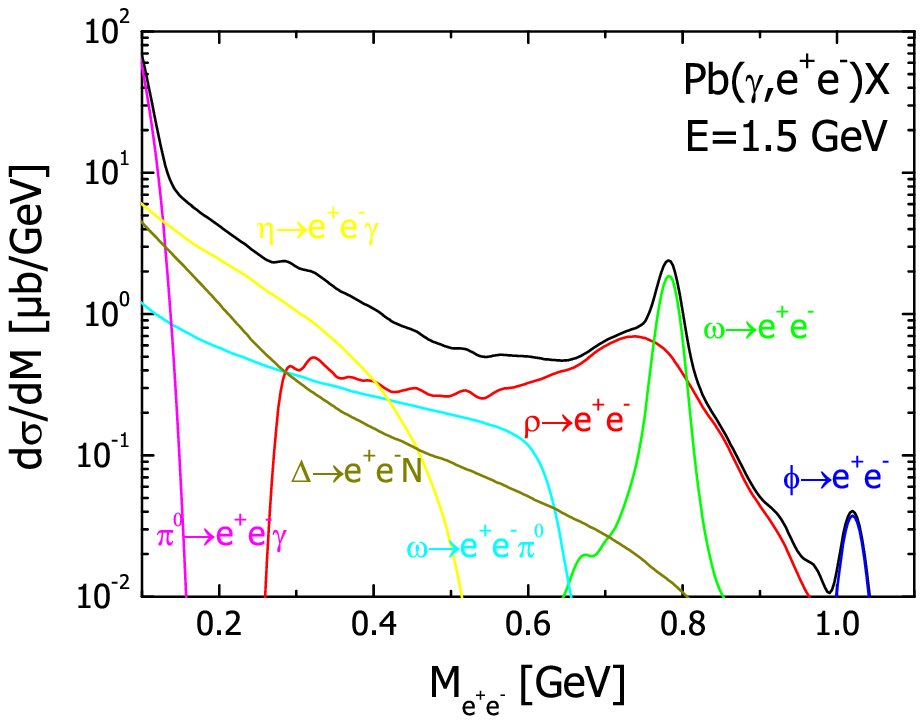,scale=0.9}}
\centerline{\epsfig{file=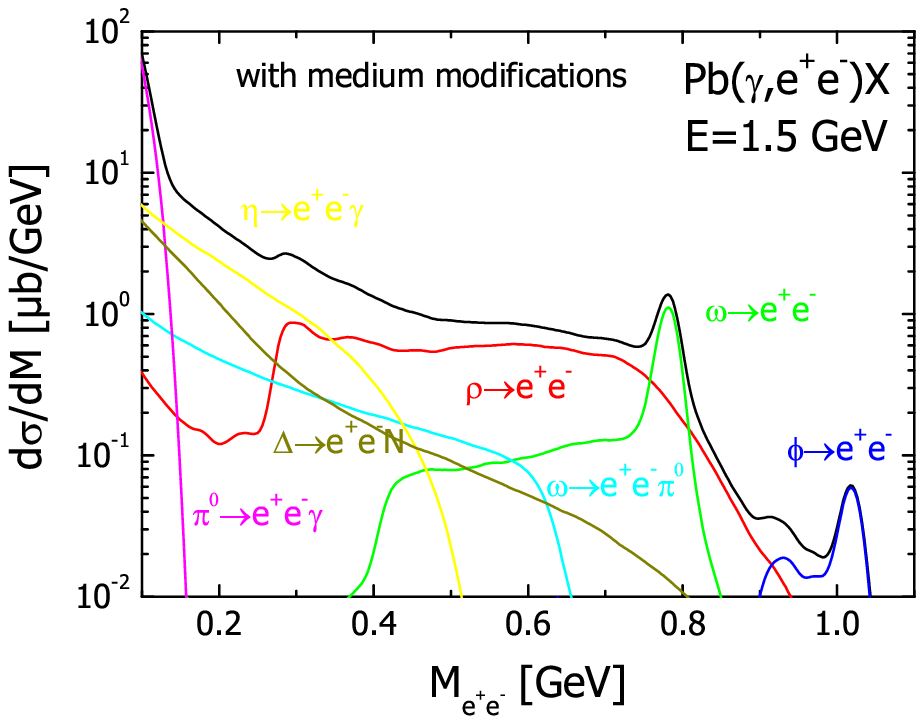,scale=0.9}}
\caption{Hadronic contributions to dilepton invariant mass spectra
for $\gamma + ^{208}Pb$ at 1.5 GeV photon energy). Indicated are the
individual contributions to the total yield; compare with
Fig.~\ref{CERES} (from \protect\cite{Muehlichpriv}).}
\label{Fige+e-}
\end{figure}
In \cite{Effephot} we have analyzed the photoproduction of dileptons
on nuclei in great detail. After removing the Bethe-Heitler
contribution the dilepton mass spectrum in a 2 GeV photon-induced
reaction looks very similar to that obtained in an ultrarelativistic
heavy-ion collision (Fig.\ \ref{CERES}). The radiation sources are
all the same in both otherwise quite different reactions. The
photon-induced reaction can thus  be used as a baseline experiment
that allows one to check crucial input into the simulations of more
complicated heavy-ion collision.

A typical result of such a calculation for the dilepton yield --
after removing the Bethe-Heitler component -- is given in Fig.\
\ref{Fige+e-}. The lower part of Fig.\ \ref{Fige+e-} shows that we
can expect observable effect of possible in-medium changes of the
vector meson spectral functions in medium on the low-mass side of
the $\omega$ peak. In \cite{Effephot} we have shown that these
effects can be drastically enhanced if proper kinematic cuts are
introduced that tend to enhance the in-medium decay of the vector
mesons. There it was shown that in the heavy nucleus $Pb$ the
$\omega$-peak completely disappears from the spectrum if in-medium
changes of width and mass are taken into account. The sensitivity of
such reactions is thus as large as that observed in
ultrarelativistic heavy-ion reactions.

An experimental verification of this prediction would be a major
step forward in our understanding of in-medium changes. The ongoing
g7 experiment at JLAB is presently analyzing such data
\cite{Weygand}. This experiment can also yield important information
on the time-like electromagnetic formfactor of the proton and its
resonances \cite{MoselHirsch95} on which little or nothing is known.

\section{Conclusions}\label{concl}
In this talk I have first outlined the theoretical motivation for
studies of in-medium properties of hadrons and their relation to
QCD. I have then shown that transport theory can be used to
calculate the in-medium properties of hadrons and their interactions
in a consistent way.  In particular photonuclear reactions on nuclei
can give observable consequences of in-medium changes of hadrons
that are as big as those expected in heavy-ion collisions which
reach much higher densities, but proceed farther away from
equilibrium. Special emphasis was put in these lectures not so much
on the theoretical calculations of hadronic in-medium properties
under simplified conditions, but more on the final, observable
effects of any such properties. I have discussed that for reliable
predictions of observables one has to take the final state
interactions with all their complications in a coupled channel
calculation into account; simple Glauber-type descriptions are not
sufficient.

A first, well known example for the effects of in-medium
interactions is given by the spectral functions of nucleons inside
nuclei.  As an example that is free from complications by FSI I have
shown that in photonuclear reactions in the 1 - 2 GeV range the
expected sensitivity of dilepton spectra to changes of the $\rho$-
and $\omega$ meson properties in medium is as large as that in
ultrarelativistic heavy-ion collisions and that exactly the same
sources contribute to the dilepton yield in both experiments. While
the dilepton decay channel is free from hadronic final state
interactions this is not so when the signal has to be reconstructed
from hadrons present in the final state. Nevertheless, the $\omega$
photoproduction, identified by the semi-hadronic $\pi^0 \gamma$
decay channel, seems to exhibit a rather clean in-medium signal.

\section*{Acknowledgement}

I gratefully acknowledge many stimulating discussions with J. Lehr,
H. Lenske, S. Leupold and P. Muehlich; many of the results discussed
in these lectures are based on their work. This work has been
supported by the Deutsche Forschungsgemeinschaft, partly through the
SFB/Transregio 16 ``Subnuclear Structure of Matter''.

\end{document}